\documentstyle[prb,aps,twocolumn,graphicx,amssymb]{revtex}
 
\newcommand{\ud}{\,{\mathrm d}}
\newcommand{\uheq}{h_{\mathrm{eq}}}
\newcommand{\uhan}{h_{\mathrm{an}}}
\newcommand{\uaeq}{a_{\mathrm{eq}}}
\newcommand{\uaan}{a_{\mathrm{an}}}

\begin{document}
\draft
\title{Evolution and stability of a magnetic vortex in small
  cylindrical ferromagnetic particle under applied field}
\author{Konstantin Yu. Guslienko\cite{address}}

\address{Korea Institute for Advanced Study, 207-43
  Cheongryangri-dong, Dongdaemun-gu, Seoul 130-012 Korea}

\author{Konstantin L. Metlov\cite{email}}

\address{Institute of Physics ASCR, Na Slovance 2, Prague 8, CZ-18221}
\date{\today}
\maketitle

\begin{abstract}
  The energy of a displaced magnetic vortex in a cylindrical particle
  made of isotropic ferromagnetic material (magnetic dot) is
  calculated taking into account the magnetic dipolar and the exchange
  interactions. Under the simplifying assumption of small dot
  thickness the closed-form expressions for the dot energy is written
  in a non-perturbative way as a function of the coordinate of the
  vortex center. Then, the process of losing the stability of the
  vortex under the influence of the externally applied magnetic field is
  considered. The field destabilizing the vortex as well as the field
  when the vortex energy is equal to the energy of a uniformly
  magnetized state are calculated and presented as a function of dot
  geometry. The results (containing no adjustable parameters) are
  compared to the recent experiment and are in good agreement.
\end{abstract}

\paragraph*{Introduction.} The unusual magnetic properties of
sub-micron cylindrical magnetic particles and their periodical
two-dimensional arrays drawn much attention because of their possible
potential as a magnetic storage as well as an interesting model system
for studying of the magnetization reversal
\cite{MFRDC97,WTKPWC98,C00,PSN00}. The applications of such patterned
magnetic films for magnetic information storage (MRAM, for instance)
are promising.

In this work the isolated magnetic particles made of isotropic (soft)
magnetic material shaped as circular cylinders (referred hereafter as
``dots'') are considered. The magnetic properties of the dots are
governed by the magnetic dipolar and the exchange interactions. When
the external magnetic field is absent, three parameters of the
dimension of length define completely the magnetic structure of a dot.
These parameters are: the dot radius $R$, the dot thickness $L$ and
the exchange length $L_E=\sqrt{C/M_S^2}$, where $C$ is the exchange
constant and $M_S$ is the saturation magnetization of the material.
The phase diagram of small $L/L_E \lesssim 4$, $R/L_E \lesssim 4$ dots
in no applied magnetic field showing the range of dot parameters where
the magnetic vortex (curl) is the ground state was calculated by Usov
and co-workers\cite{UP93,UP94}. Vortex states were indeed observed in
polycrystalline $Co$ \cite{FC00} and $FeNi$
\cite{PSN00,CKAWT99,SOHSO00,RPSSZW00,SHZ00} cylindrical dots.

So far there is no theory of magnetization reversal of a dot in a
vortex state when the magnetic field $h=H/4 \pi M_S$ applied in the
direction perpendicular to the cylinder axis. This process taking
place via nucleation, displacement and annihilation of a single
magnetic vortex\cite{SHZ00} is considered in this work. There are
three characteristic fields describing this process: the ``vortex''
nucleation $h_n$ and annihilation $\uhan$ critical fields (for
transition to/from the curling state from/to the saturated state) and
also the field $\uheq$, corresponding to the vortex state having the
same energy as the uniform one.  The size dependent fields of vortex
equilibrium $\uheq$ and annihilation $\uhan$ are calculated in this
paper. The field of vortex nucleation $h_n$ is determined by the
imperfections of the dot shape (and/or material) and was
experimentally found to have nearly no correlation to the dot geometry
\cite{SHZ00}. The field $\uheq$ enters directly the phase diagram in
coordinates $h$, $L/L_E$, $R/L_E$ whose zero field part\cite{UP94} was
already presented.

A realistic description of the non-uniform magnetization distribution
in an isolated dot at zero applied field was already considered:
first, with the help of guessed trial functions\cite{A90} and then
using the variational principle\cite{UP93}. These approaches heavily
rely on the cylindrical coordinate system and are hard to generalize
for the case when the vortex is deformed by the in-plane magnetic
field.  Recently, another variational method to construct the
magnetization distributions in small exchange-dominated cylinders (not
limited to circular ones) was presented \cite{M01_solitons}. It allows
to obtain the structure of a displaced vortex, which is a starting
point for the calculations done in this work.

\begin{figure}
  \begin{center}
    \includegraphics[scale=0.45]{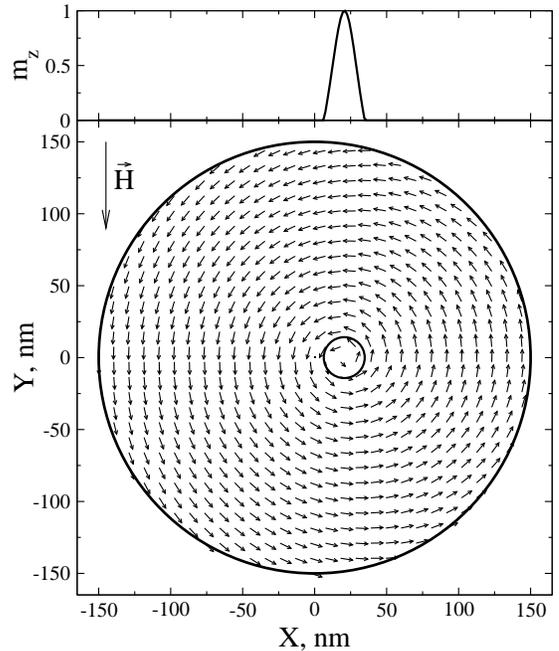}
    \caption{The magnetization distribution (\ref{eq:magn_conformal}) 
      with the equilibrium values $a\approx0.14$ and $c\approx0.092$
      corresponding to a permalloy ($L_E=18\,nm$) cylinder with
      $R=150\,nm$, $L=15\,nm$ in the magnetic field $h=H/(4 \pi
      M_S)=-0.025$ applied parallel to the cylinder $0Y$ axis. The
      out-of-plane magnetization component on the line $Y=0$ is shown
      on the upper plot.}
    \label{fig:magnetization_distribution}
  \end{center}
\end{figure}

\paragraph*{Equilibrium parameters and stability of displaced vortex.}
In a thin ferromagnetic cylinder, with the thickness less or of the
order of the exchange length, the magnetization distribution is
constant along the thickness and is essentially two-dimensional. In
the Cartesian coordinate system with coordinates $\{X,Y,Z\}$, chosen
in such a way that the axis $0Z$ is parallel to the cylinder axis, the
normalized magnetization vector field
$\vec{m}(\vec{r})=\vec{M}(\vec{r})/M_S$, $|\vec{m}|=1$ is independent
on the coordinate $Z$, so that $\vec{r}=\{X,Y\}$. In this case the
magnetization distribution corresponding to the minimum of the
exchange energy functional
\begin{equation}
  \label{eq:exch_functional}
  E_{ex}=\frac{1}{2} C L \int \ud^2 \vec{r} 
  \sum_{i=X,Y,Z} (\vec{\nabla} m_i(\vec{r}))^2
\end{equation}
and having the structure of a displaced magnetic vortex
is\cite{M01_solitons}
\begin{equation}
  \label{eq:magn_conformal}
  w(z,\overline{z})= \left\{
    \begin{array}{ll}
      \frac{\imath}{c}\frac{z-a}{1-\overline{a}z} &
      |z-a| < c |1-\overline{a}z| \\
      \imath \frac{(z-a)(1- a \overline{z})}{|z-a| |1-\overline{a}z|} &
      |z-a| \geq c |1-\overline{a}z|
    \end{array}
  \right. ,
\end{equation}
where the components of the magnetization vector are expressed as
$m_x+\imath m_y = 2 w/(1+w\overline{w})$ and
$m_z=(1-w\overline{w})/(1+w\overline{w})$ through the complex function
$w(z,\overline{z})$ of the complex variable $z=X/R+\imath Y/R$, the
horizontal line over a variable denotes the complex conjugation
$\overline{z} = X/R - \imath Y/R$, the straight brackets stand for the
absolute value of a complex number $|z|^2 =z\overline{z}$.  The
expression (\ref{eq:magn_conformal}) gives the magnetization
distribution minimizing the functional (\ref{eq:exch_functional}) for
all values of a scalar parameter $c$, related to the radius of the
vortex core and a complex parameter $a=X_C/R + \imath Y_C/R$,
describing the position of the vortex center. The case of $a=0$
corresponds to the Usov's vortex solution\cite{UP93} which satisfies
the conditions $(\vec{m} \cdot \vec{n})_S=0$ and $(\partial \vec{m} /
\partial\vec{n})_S=0$ on the side dot boundary. The conformal mapping
$z \Rightarrow (z-a)/(1-\overline{a}z)$ used\cite{M01_solitons} to
obtain (2) holds the boundary condition $(\partial \vec{m} /
\partial\vec{n})_S=0$ unchanged, whereas the side magnetic charges
$(\vec{m} \cdot \vec{n})_S$ appear.  If the magnetic field is applied
along $Y$ axis it is clear from the symmetry that the vortex will be
displaced parallel to the $X$ axis, which allows to assume that the
parameter $a$ is real. The expression (\ref{eq:magn_conformal}) yields
correct limiting cases $a=0$ (centered vortex) and $a=1$
(single-domain or saturated dot) and is applicable at small enough $R$
and $L$.  The example distribution at an intermediate value of $a$ is
given in Fig.~\ref{fig:magnetization_distribution}. To find particular
values of the parameters $a$ and $c$ corresponding to the given dot
geometry, material and applied magnetic field it is necessary to
include into consideration the magnetostatic and Zeeman energies.

The magnetostatic energy has two surface (on the dot faces and sides)
and a negligible (for small dot thickness) volume contributions. To
simplify the further consideration we shall assume that the dot thickness
is small enough to neglect the volume magnetostatic contribution.

In the further calculations two polar coordinate systems with the same
notation for coordinates $r$, $\phi$ will be used. One of the
coordinate systems will be centered in the dot center and have the
dimensionless polar radius $\rho=r/R$. The other coordinate system
centered at the vortex core center is also polar, because the vortex
core (the region of the dot with non-zero $m_z$) has exactly the
circular boundary for all values of $a$. The center of the vortex core
(note the difference with the vortex center where $m_z=1$) is situated at the
coordinates $X_V=a(1-c^2)/(1-a^2c^2)$, $Y_V=0$ for real $a$, and its
radius $R_V=c(1-a^2)/(1-a^2c^2)$. The dimensionless polar radius for
the coordinate system centered at the vortex core is $\rho=r/R_V$.

The density of magnetic charges produced by the distribution
(\ref{eq:magn_conformal}) on the faces of the dot in the polar
coordinate system centered at the vortex core is
\begin{equation}
  \label{eq:face_charges}
  m_z = \sigma(\rho,\phi)=\cos \delta \frac{1-\rho^2}
  {1+\rho^2-2\rho \sin \delta \cos \phi}, 
\end{equation}
with $\delta  = 2 \arctan (a c)$.
The magnetostatic energy of these charges normalized to $4 \pi M_S^2
V$ ($V= L \pi R^2$ is the dot volume, all energies further in the text
will be given in this normalization) is
\begin{equation}
  \label{eq:face_energy1}
  e^{\mathrm{face}}_{MS}=
  \frac{E^{\mathrm{face}}_{MS}}{4 \pi M_S^2 V}=
  \frac{R_V^3}{L R^2} \left( G(0) - G(L/R_V) \right),
\end{equation}
where 
\begin{equation}
  \label{eq:face_g}
  G(x) =
  \!\!\int\!\!
  \frac{\sigma(\rho_1,\phi_1) \sigma(\rho_2,\phi_2) 
    \rho_1 \ud \rho_1 \rho_2 \ud \rho_2 \ud \phi_1 \ud \phi_2}
  {(2\pi)^2\sqrt{x^2 + \rho_1^2 + \rho_2^2 - 2 \rho_1 \rho_2 \cos (\phi_1-\phi_2)}},
\end{equation}
the integration runs from $0$ to $1$ in $\rho_1$, $\rho_2$ and from
$0$ to $2\pi$ in $\phi_1$, $\phi_2$. Assuming that the parameter $ac$
is small (which is justified by our numerical calculations, also see
the caption to Fig.~\ref{fig:magnetization_distribution}) the last
expression in zero-th order in $ac$ is
\begin{equation}
  \label{eq:face_g0}
    e^{\mathrm{face}}_{MS}=\frac{c^3(1-a^2)^3}{g}\left(G_U\left(0\right) - 
      G_U(\frac{g}{c(1-a^2)}) \right)
\end{equation}
where $G_U(x)=G(x, \delta \rightarrow 0)$ is the magnetostatic
function of non-displaced vortex core\cite{UP93}
and $g=L/R$.  While this expression is approximate it captures the
details of the exact one (\ref{eq:face_energy1}) well, and has error
not more than $1\%$ as we have checked numerically.

The surface magnetic charges on the sides of the dot are
\begin{equation}
  \label{eq:side_charges}
  (\vec{m}\vec{n})_S= \sigma(\phi) = 2 a \sin \phi \frac{a \cos \phi - 1}
  {1 + a^2 - 2 a \cos \phi},
\end{equation}
which is written in the cylindrical coordinate system $r$, $\phi$ centered at
the dot center. The energy of these charges can be expressed as
\begin{equation}
  \label{eq:side_energy1}
  e^{\mathrm{side}}_{MS}=
  \!\!\int\!\!
  \frac{\sigma(\phi_1)\sigma(\phi_2) \ud z_1 \ud z_2 \ud \phi_1 \ud \phi_2}
  {2 g (2\pi)^2 \sqrt{2(1-\cos(\phi_1-\phi_2))+(z_1-z_2)^2}},
\end{equation}
where the integration runs from $0$ to $g$ in $z_1$,$z_2$ and from $0$
to $2\pi$ in $\phi_1$, $\phi_2$. The last expression can be simplified
using the summation theorem for the Bessel's functions into
\begin{eqnarray}
  \label{eq:side_energy2}
  e^{\mathrm{side}}_{MS} & = &
  \frac{1}{2} \left[ a^2 (a^2-2)^2 I_1(g) +
    (1-a^2)^2 \sum_{\mu=1}^\infty a^{2\mu} I_\mu(g) \right] \\
  I_\mu(g) & = & \int_0^\infty \frac{\ud k}{k} 
  \left(1-\frac{1-\exp(-k g)}{k g} \right)
  J_\mu^2(k),
\end{eqnarray}
where  $J_\mu(x)$ is the Bessel's function of the first kind and order $\mu$.
Minimizing the total energy we have kept only one term in the above
series, which was sufficient to have the error less than $3\%$.

The total magnetization of the dot (having only one non-zero component along
$Y$ axis), which enters the Zeeman energy is
\begin{eqnarray}
  \label{eq:magn_density}
  \langle m_y \rangle \! = 
  \!\!\int\!\!
    \frac{((1+a^2) \rho \cos \phi - a (1+\rho^2)) \rho \ud \rho \ud \phi }
    {\sqrt{a^2+\rho^2 - 2 a \rho \cos \phi}
    \sqrt{1+ a^2 \rho^2 - 2 a \rho \cos \phi}},
\end{eqnarray}
where the integration from $0$ to $1$ in $\rho$ and from $0$ to $2\pi$
in $\phi$ is performed in the coordinate system centered on the dot. This
integral can be approximated by $\langle m_y \rangle = a(15 a^4 + 34
a^2 - 193)/144$ with the error less than $1\%$ according to our
numerical calculations.

The exchange energy can be evaluated using its
expression\cite{M01_solitons} directly through the function
$w(z,\overline{z})$, or, alternatively, using
(\ref{eq:exch_functional}) as
\begin{equation}
  \label{eq:exch}
  e_{\mathrm{EX}}= 
  \frac{1}{4\pi}\left(\frac{L_E}{R}\right)^2 (2 - \log (c) ),
\end{equation}
which is independent on $a$.

The total energy density of the particle is $e=e_{\mathrm{ex}}+
e^{\mathrm{face}}_{MS}+ e^{\mathrm{side}}_{MS} - h\langle m_y \rangle$
and should be minimized with respect to the parameters $a$ and $c$. These
parameters are coupled through the $e^{\mathrm{face}}_{MS}$ term which
depends on both of them. However, if we use $q=c(1-a^2)/g$ instead of
$c$ the parameters $a$ and $q$ become decoupled and the energy can be
minimized with respect to each independently.

The only terms depending on $q$ are
$e_{\mathrm{EX}}+e^{\mathrm{face}}_{MS} = \lambda^2 g^2 (2 - \log q g)
+ q^3 g^2 Q(1/q) + const$, where
$Q(x)=G_U(0)-G_U(x)$, $\lambda=L_E/(L\sqrt{4\pi})$. The equilibrium
value of $q$, then, satisfies the following equation
\begin{equation}
  \label{eq:equilbr_q}
  q \frac{\ud}{\ud q} \left[q^3 Q(1/q)\right] = \lambda^2,
\end{equation}
which is exactly the equation\cite{UP93,UP94} for the equilibrium
radius of the vortex core at zero field.

The rest of terms in the total energy $e_1(a,
h)=e^{\mathrm{side}}_{MS} - h \langle m_y \rangle + \lambda^2 g^2 \log
(1-a^2)$ depend only on $a$, the last term comes from the exchange
energy after changing the independent variable from $c$ to $q$.

\begin{figure}
  \begin{center}
    \includegraphics[scale=0.4]{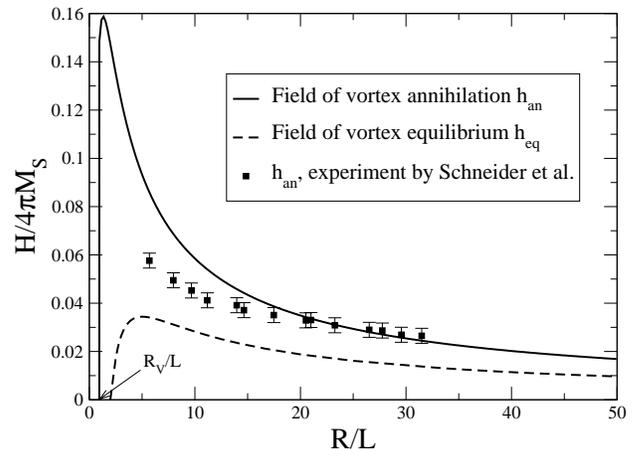}
    \caption{
      Characteristic fields of the permalloy $L_E=18\,nm$ dots of the
      thickness $L=15\,nm$ as a function of their radius. The experimental
      points are taken from the literature\cite{SHZ00}.}
    \label{fig:cylinder_energy}
  \end{center}
\end{figure}

The vortex loses its stability when at a certain value of the external
field $h=\uhan$ the energy minimum it is in turns to a saddle point or a
maximum. This leads to the following system of equations for
$\uaan$, $\uhan$ :
\begin{equation}
  \label{eq:stability_field}
  \left\{
    \begin{array}{l}
    \partial e_1(\uaan,\uhan)/
    \partial \uaan=0 \\
    \partial^2 e_1(\uaan,\uhan)/
    \partial \uaan^2=0.
    \end{array}
  \right.
\end{equation}
These equations were solved numerically for $L_E/L=18\,nm/15\,nm$
(corresponding to the experiment\cite{SHZ00}) and the resulting
dependence of the vortex annihilation field $\uhan$ on the reduced dot
radius $1/g=R/L$ is shown in Fig.~\ref{fig:cylinder_energy}.  The
resulting $\uhan(1/g)$ curve agrees well with the experimental
points\cite{SHZ00}.

The field $\uheq$ when the energy of the displaced vortex in the
external magnetic field is equal to the energy of the uniformly
magnetized state was determined from the equations:
\begin{equation}
  \label{eq:equilib_field}
  \left\{
    \begin{array}{l}
    \partial e_1(\uaeq,\uheq)/
    \partial \uaeq=0 \\
    e_1(\uaeq,\uheq) = 
    e^{\mathrm{side}}_{MS} (a\rightarrow 1) - \uheq,
    \end{array}
  \right.
\end{equation}
and also plotted in Fig.~\ref{fig:cylinder_energy}. This expression
takes into account the fact that the exchange and the face
magnetostatic energies are equal to zero for the single-domain
(saturated in-plane) dot.

\paragraph*{Discussion.}
The calculations of the annihilation field explain reasonably the
experiment over a wide range of $1/g$. The differences between the
measured and calculated $\uhan$ curves at small $R/L<10$ can be the
result of increasing role of thermal fluctuations in small particles.
The peak on the calculated dependencies of both $\uhan$ and $\uheq$ on
$R$ (see Fig.~\ref{fig:cylinder_energy}) due to their decrease with
the cylinder radius at small $R$ is not present in the experimental
data\cite{SHZ00} and needs an explanation.

It can be explained as the result of inherent (not requiring the
applied field) instability of vortex in the cylinders of small radius.
There are two kinds of forces acting on the vortex in a cylinder, the
stabilizing (due to the energy terms increasing as the vortex comes
off the dot center) and destabilizing ones. In the case considered in
this paper the main stabilizing force is produced by the energy of
magnetic charges on the cylinder side. The rate of the increase of
this energy term (\ref{eq:side_energy2}) with $a$ (the actual force)
decreases with the dot radius $R$ at a fixed $L$. This decrease is
related to the decrease in the possible amount of magnetic charges
forming (the area of side decreases with $R$) and also to the
decreasing distance between the positive and negative ones, whose
negative interaction energy strengthens as a result.

The destabilizing force is due to the energy of the magnetic charges
on the cylinder faces (\ref{eq:face_energy1}), which is positive for
the centered vortex and is zero when the vortex is completely out of the
cylinder. This force is independent on the dot radius as long as the
vortex core fits completely within the particle.

Therefore, as the radius of the cylinder decreases there is a point
(called the stability radius $R_S$) where the destabilizing force
starts to outweigh the stabilizing one and the vortex becomes unstable
in such a cylinder. If the cylinder radius is smaller than $R_S$ {\em no
  magnetic field is required} to destabilize the vortex and $\uhan=0$.
This means that the field $\uhan$ decreases at small cylinder radii to
become zero at $R=R_S$.

The similar decrease of $\uheq$ happens because the smaller is the
cylinder the smaller is the energy of the uniformly magnetized state
in it and it is closer to the energy of the cylinder with a vortex.
Until, at the so-called single domain radius $R_C$ these energies
become equal with no applied field. This means at small radii $\uheq$
also decreases with the radius until it is exactly zero at $R=R_C$.

The values of both $R_S$ and $R_C$ have sense only when they are
larger then the vortex core radius $R_V$. This is not the case for the
cylinder thickness $L/L_E=15/18$ shown in
Fig.~\ref{fig:cylinder_energy}, where $R_S<R_V$). Our calculations
show that at smaller dot thicknesses the case when $R_S, R_C > R_V$
can be realized and $\uhan$, $\uheq$ curves start from zero at small
radii.

As we have mentioned in the introduction, the dependence of $R_C/L_E$
on $L/L_E$ was already reported in literature\cite{UP94}, its
calculation does not require a model of displaced vortex. The
coincidence of the cylinder radius at which $\uheq$ turns to zero with
$R_C$ was one of the checks we did to verify the results.

The approximations we performed were focused on calculating the dot
energy at arbitrarily large vortex displacements, that is, not to
perform a perturbative expansion limiting us to a particular range of
displacements. However, for the calculation of $R_S$ corresponding to
the vortex displacement mode given by Eq. (\ref{eq:magn_conformal})
the expansion at $a=0$ up to second order is enough. The
approximations done here, especially Eq. (\ref{eq:face_g0}), are,
strictly speaking, not valid up to the second order in $a$ at $a=0$.
Therefore, the stability radius resulting from this calculation should
be taken with caution. The mentioned expansion is easy to perform and
we plan to present the precise estimate of $R_S$ for the mode
(\ref{eq:magn_conformal}) in one of our forthcoming papers.

Some correlation of the vortex annihilation field with the dot average
in-plane demagnetization factor in saturated state was already
noted\cite{FC00}. That can be explained as follows. The vortex shift
$a$ in the external magnetic field is determined by competition of the
side magnetostatic and Zeeman energies. The main contribution in Eq.
(\ref{eq:side_energy2}) to the magnetostatic energy from the dot
boundary is proportional to $I_1(g)$ being exactly the single-domain
dot in-plane average demagnetization factor\cite{GCS00}. But this is
correct only for large enough dot radius (considerably above $R_S$,
$R_C$) when the face magnetostatic energy and exchange energy can be
neglected.

\paragraph*{Conclusions.}
In summary, the size-dependent vortex equilibrium and annihilation
fields have been studied theoretically for circular cylindrical
magnetic dots. Simple analytical approximation suggested for the dot
magnetization distribution allowed us to consider the transition from
the vortex to single-domain dot magnetization state under applied
magnetic field. The calculation results agree well with the
experimental data within the limit of weak inter-dot coupling. Further
studies of static and dynamic properties of magnetization in isolated
and magnetostatically coupled dots are in progress.

The support by Korea Institute for Advanced Study is gratefully
acknowledged. The work was supported in part by the Grant Agency of
the Czech Republic under projects 202/99/P052 and 101/99/1662.

%

\end{document}